# Evidence for Neutrino Oscillations I: Solar and Reactor Neutrinos

A. B. McDonald

SNO Institute, Physics Department, Queen's University, Kingston, Canada, K7L 3N6

This paper discusses evidence for neutrino oscillations obtained from measurements with solar neutrinos and reactor neutrinos.

## 1. INTRODUCTION

After many years of experimental effort, evidence has now been obtained for neutrino oscillations in several areas of study. The present paper will review recent evidence in measurements of solar neutrinos and reactor neutrinos, the following paper will address atmospheric neutrinos and accelerator neutrinos.

Since the pioneering solar neutrino measurements of Ray Davis and co-workers [1] too few neutrinos were observed in several experiments that were sensitive exclusively or primarily to electron neutrinos. Recent measurements of $^8$B solar neutrinos by the Sudbury Neutrino Observatory [2,3,4,5] observed electron neutrinos through the charged current reaction on deuterium and all active neutrino types through the neutral current reaction. Clear evidence was obtained for the conversion of about 2/3 of the electron neutrinos to other active neutrino types. The flux of all active neutrinos was in close agreement with solar model calculations [6] of the flux of $^8$B neutrinos produced in the Sun, providing strong confirmation of those models. Following these observations, the KamLAND experiment [7] observed a deficit of electron anti-neutrinos produced by Japanese power reactors, confirming the flavor change with neutrino oscillation parameters very similar to those extracted from fits to the set of solar neutrino measurements to date.

These measurements for solar and reactor are in agreement with models for the oscillation of neutrinos with flavor eigenstates related to mass eigenstates via a mixing matrix referred to as the Maki-Nakagawa-Sakata-Pontecorvo (MNSP) [8] matrix. The solar neutrinos undergo matter-induced oscillations via the Mikheyev-Smirnov-Wolfenstein (MSW) [9] mechanism in the Sun. The combination of solar and reactor measurements can be used to relegate a number of other flavor change mechanisms to small sub-dominant status, including oscillation to sterile neutrinos, resonant spin-flavor oscillation and neutrino decay.

The format of the paper is as follows. First, solar neutrino measurements to date will be discussed, including the evidence for neutrino flavor change. Secondly, the evidence for disappearance of reactor anti-neutrinos will be discussed and then the full set of evidence will be discussed in terms of the MNSP and MSW theories. The observation of flavor change and interpretation in terms of neutrino oscillation is clear evidence for new physics beyond the Standard Model for elementary particles. This provides a strong motivation for further measurements that seek to define the oscillation parameters more accurately, with the hope of defining the correct theoretical extensions of the Standard Model. Plans for future measurements of solar and reactor neutrinos are discussed to show the prospects for improving the accuracy of parameters in the mixing matrix and studying other possible sub-dominant processes.

## 2. SOLAR NEUTRINOS

### 2.1. Solar Neutrino Measurements prior to 2001

Beginning with the pioneering measurements of Ray Davis and collaborators [1] in the 1960's, too few solar neutrinos were observed compared to solar model calculations by another pioneer of the field, John Bahcall, and others. The spectrum of neutrinos obtained by the latest of such calculations [6] is shown in Figure 1, together with uncertainties in the calculated fluxes and thresholds for neutrino detection by various experiments to date.

Davis' radiochemical measurements with Chlorine were sensitive primarily to $^7$Be and $^8$B neutrinos. The detector was operated from 1968 until 2000 and the cumulative average of the data [1] is 2.56±0.16(stat)±0.16(syst) SNU, where a Solar Neutrino Unit (SNU) is defined as one electron neutrino capture per $10^{+36}$ atoms of $^{37}$Cl per second. This result is substantially smaller than the predictions of Standard Solar Model calculations [6] of 8.5+1.8−1.8 SNU [11]. The measurements made by the SAGE [10], GALLEX [11] and GNO [12] experiments with Ga as a target material were primarily sensitive to the lowest energy pp neutrinos, for which the total flux is strongly constrained by solar luminosity. The results from these experiments to date are as follows: GALLEX + GNO: 70.8 ± 4.5(statistical) ± 3.8(systematic) SNU, SAGE: $70.9^{+5.3}_{-5.2}(\text{stat})^{+3.7}_{-3.2}(\text{syst})$ SNU. These numbers are in excellent agreement and are much smaller than the predictions of the standard solar model (131 SNU) [6].

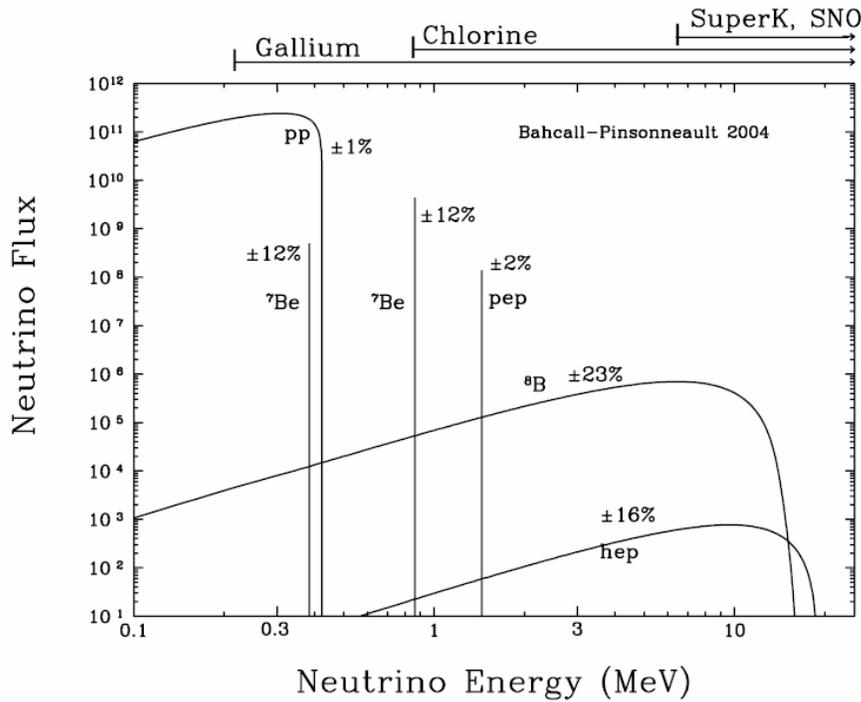

Figure 1 Neutrino fluxes from solar model calculations [6]. Thresholds for experiments are indicated at the top of the figure.

The Kamiokande detector [13] with 1000 tons of light water operated from 1986 to 1993. The water was subsequently replaced by liquid scintillator to create the KamLAND detector. The Super-Kamiokande [14] detector began operation in 1996 with a 22.5 kton fiducial volume of light water. Elastic scattering of neutrinos from electrons was used to observe solar neutrinos with thresholds as low as 5 MeV neutrino energy. This reaction is primarily sensitive to electron neutrinos, which have a cross section about six times larger than mu or tau neutrinos. With the thresholds used, the experiments were largely sensitive to solar neutrinos from $^8$B decay, with a negligible contribution from the hep reaction.

For Kamiokande, the flux of solar neutrinos from $^8$B decay was measured [13] to be $2.8 \pm 0.19(\text{stat}) \pm 0.33(\text{syst}) \times 10^{+6}$ cm$^{-2}$ s$^{-1}$, clearly less than the solar model prediction [6] of $5.82 (1 \pm 0.23) \times 10^{+6}$ cm$^{-2}$ s$^{-1}$. The flux measured by Super-Kamiokande [14] is $2.35 \pm 0.02(\text{stat}) \pm 0.08(\text{syst}) \times 10^{+6}$ cm$^{-2}$ s$^{-1}$, and is also significantly smaller than the solar model prediction.

Although the measurements with Cl, Ga and light water all showed deficits in neutrino flux, they had different sensitivities to solar fluxes from the various reactions in the Sun. There were many attempts to identify solar effects that could provide such a reduction in the fluxes but there was no conclusive effect identified in the models that could provide a full explanation. At the same time, solar models provided excellent agreement with other solar properties, including the results of helioseismology measurements. [15]

Alternative explanations for the flux deficits were proposed, with neutrino oscillations arising from neutrino eigenstates with finite mass as the leading explanation. However, it was difficult to be certain that oscillations were taking place and that the deficits did not arise from uncertainties in the solar models.

The primary candidate to explain the observed solar neutrino deficit through a change in neutrino properties was neutrino oscillation. Such oscillation can occur if flavor eigenstates for the three active neutrino types (l = e, μ, τ) are related to mass eigenstates (i) via the MNSP [8] mixing matrix $U_{li}$:

$$|v_l\rangle = \Sigma U_{li}|v_i\rangle.$$

The MNSP matrix $U_{li}$ can be written for three active neutrino types as:

$$U_{li} = \begin{pmatrix} c_{12} & s_{12} & 0 \\ -s_{12} & c_{12} & 0 \\ 0 & 0 & 1 \end{pmatrix} \cdot \begin{pmatrix} 1 & 0 & 0 \\ 0 & c_{23} & s_{23} \\ 0 & -s_{23} & c_{23} \end{pmatrix} \cdot \begin{pmatrix} 1 & 0 & 0 \\ 0 & 1 & 0 \\ 0 & 0 & e^{-i\delta} \end{pmatrix} \cdot \begin{pmatrix} c_{13} & 0 & s_{13} \\ 0 & 1 & 0 \\ -s_{13} & 0 & c_{13} \end{pmatrix}$$

where $c_{ij} = \cos\theta_{ij}$, and $s_{ij} = \sin\theta_{ij}$

For non-degenerate mass eigenstates, and for small $\theta_{13}$, oscillations of solar neutrinos are dominated by the first sub-matrix involving $\theta_{12}$. The second sub-matrix dominates the oscillation of atmospheric neutrinos, the third sub-matrix involves the CP violating angle δ and the fourth sub-matrix is tested by reactor and accelerator neutrino measurements. For oscillations in vacuum in the two neutrino mixing approximation, the survival probability for solar neutrinos with energy E, that have traveled a distance L is:

$$P(v_e \rightarrow v_e) = 1 - \sin^2 2\theta_{12} \sin^2 (1.27 \frac{\Delta m_{12}^2 L}{E})$$

If the neutrinos travel in a region of high electron density, such as in regions within the Sun or Earth, the interaction of neutrinos with electrons can produce matter enhancement of the oscillation process, known as the MSW effect [9]. The MSW process could result in an energy dependence of the oscillation probability beyond the vacuum dependence and could produce a difference in the neutrino fluxes for neutrinos that have traveled through the Earth. This could produce differences in the measured fluxes at detectors for day and night time periods.

In order to seek clear evidence for effects that were independent of solar model calculations and could only arise from changes in neutrino properties, time dependence and energy distortion was also studied in the solar neutrino measurements. The Super-Kamiokande (SK) experiment had by far the best statistical accuracy for such studies. Studies by SK of the variation in flux as a function of time showed no statistically significant effects [14]. The neutrino energy spectrum was found to be very similar to that expected from $^8$B decay and no distortion due to the MSW effect [9] has been observed.

If the explanation for the solar neutrino deficit arises from neutrino oscillations, the best fit to the solar neutrino data was for a region of mass and mixing angle known as the Large Mixing Angle (LMA) region, involving matter enhancement in the Sun. Figure 2 shows the spectral shape and day-night differences observed by the Super-Kamiokande experiment, compared to the expected shape and day-night asymmetry for the LMA solution, involving very little distortion or asymmetry, as observed.

In addition to the day-night studies of neutrino flux, seasonal and annual modulations were sought by Super-Kamiokande to look for effects that could arise from resonant spin-flavor precession created by a finite neutrino magnetic field interacting with solar magnetic fields. No significant effects were observed beyond the geometrical effects arising from the eccentricity of the Earth's orbit [14].

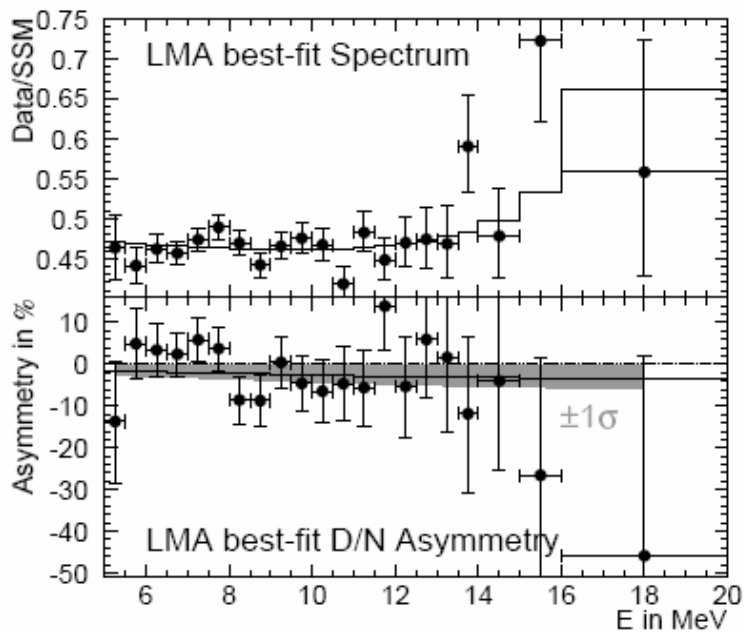

Figure 2. Solar neutrino elastic scattering energy spectrum (top) and Day/Night Asymmetry (bottom) as measured by Super-Kamiokande. The predictions (solid lines) are for the LMA solution discussed below. The gray bands are the ± σ ranges for the asymmetry over the entire energy range.

**2.2 Solar Neutrino Measurements by the Sudbury Neutrino Observatory (SNO)**

The SNO experiment was designed primarily to search for a clear indication of neutrino flavor change for solar neutrinos without relying upon solar model calculations. This was accomplished by comparing the flux of $^8$B electron neutrinos observed with the charged current reaction on deuterium with the flux of all active neutrino types observed with the neutral current reaction on deuterium. If the electron neutrino flux is smaller than the total flux this would be a clear indication of neutrino flavor change to active types.

The reactions observed in a Cerenkov detector based on 1000 tons of heavy water [16] are as follows:

$$\nu_e + d \rightarrow p + p + e^- \text{ (CC)}$$
$$\nu_x + d \rightarrow p + n + \nu_x \text{ (NC)}$$
$$\nu_x + e^- \rightarrow \nu_x + e^- \text{ (ES)}$$

The charged current (CC) reaction is sensitive exclusively to electron neutrinos, while the neutral current (NC) reaction is sensitive to all neutrino flavors (x = e, μ, τ) above the energy threshold of 2.2 MeV. The elastic scattering (ES) reaction is sensitive to all flavors as well, but with reduced sensitivity to $\nu_\mu$ and $\nu_\tau$. The CC and ES reactions are observed through the Cherenkov light produced by the electrons. The NC reaction is observed through the detection of the neutron in the final state of the reaction.

The SNO experimental plan involves three phases wherein different techniques are employed for the detection of neutrons from the NC reaction. During the first phase, with pure heavy water, neutrons were observed through the Cherenkov light produced when neutrons are captured on deuterium, producing 6.25 MeV gammas. For the second phase, about 2 tons of salt was added to the heavy water and neutron detection was enhanced through capture on Cl, with about 8.6 MeV gamma energy release and higher capture efficiency. The neutrons from the NC reaction capture on Cl and provide a more isotropic pattern at the PMT's for these events than for the CC events. This enables the fluxes from the CC and NC reactions to be extracted on a statistical basis with no constraint on the shape of the CC energy spectrum. For the third phase, the salt was removed and an array of $^3$He-filled proportional counters has been installed to provide neutron detection independent of the PMT array.

Results from the initial phase of the experiment with pure heavy water [2,3,4] showed clear evidence for neutrino flavor change, without reference to solar model calculations. The hypothesis of no flavor change was tested and found to be violated using summed data for the CC and NC reactions observed through Cerenkov processes in the heavy water. With the addition of salt it was possible to break the strong correlation between the CC and NC event types because the pattern of light on the PMT's is different for the two reactions. An event angular distribution parameter $\beta_{14}$ (defined in reference [5]) was used to separate NC and CC events on a statistical basis.

The data from the salt phase [5] are shown in Figure 3, together with the best fit to the data of the NC (8.6 MeV gamma) shape, the CC and ES reactions and a small background component determined from independent measurements.
The fluxes inferred for this fit are:

$$\Phi_{CC} = 1.59^{+0.08}_{-0.07} \text{(statistical)}^{+0.06}_{-0.08} \text{(systematic)}$$
$$\Phi_{NC} = 5.21 \pm 0.27 \text{(stat)} \pm 0.38 \text{(syst)}$$
$$\Phi_{ES} = 2.21^{+0.31}_{-0.26} \text{(stat)} \pm 0.10 \text{(syst)}$$

where these fluxes and those following in this section are quoted in units of $10^6$ cm$^{-2}$ sec$^{-1}$.

These fluxes show clearly that about two-thirds of the electron neutrinos have changed their flavor to other active neutrino types. In addition, the observed total flux of active neutrinos is in excellent agreement with the flux of $^8$B neutrinos obtained from solar models: $\Phi_{SSM} = 5.82(1 \pm 0.23)$.

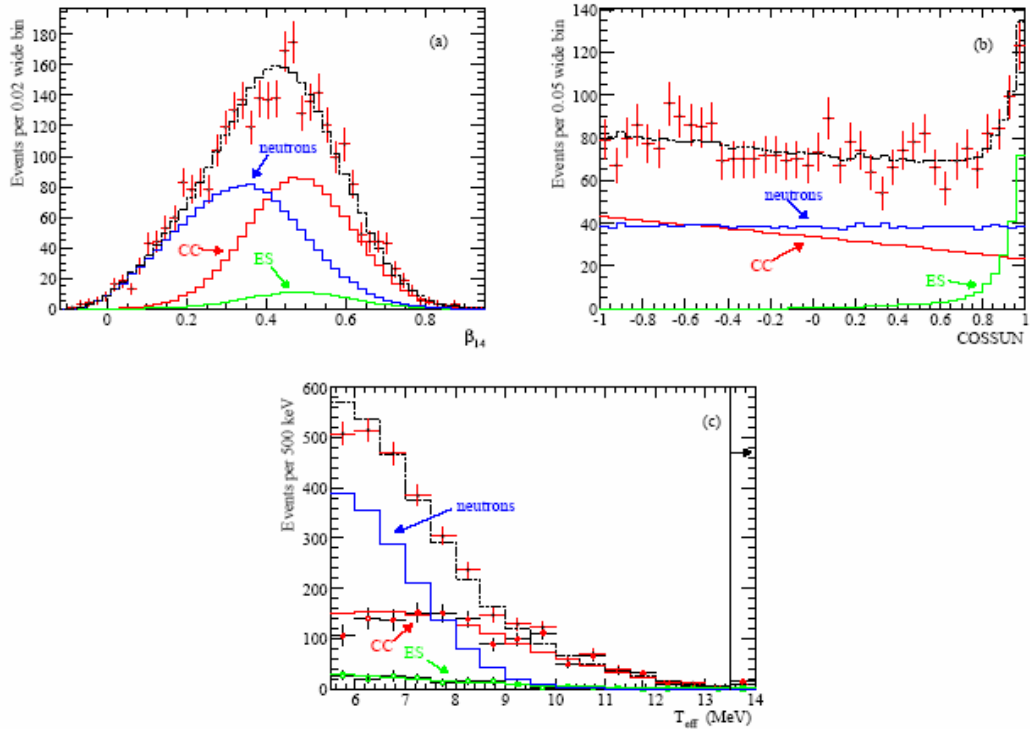

Figure 3 Distribution of (a) $\beta_{14}$, (b) cos $\theta_{sun}$ and (c) kinetic energy, for the selected events in the fiducial volume for the salt phase of SNO. The CC and ES spectra are extracted from the data using $\beta_{14}$ and cos $\theta_{sun}$ distributions in each energy bin. Also shown are the Monte Carlo predictions for CC, ES, NC + internal and external-source neutron events, all scaled to the fit results. The dashed lines represent the summed components. All distributions are for events with effective energy > 5.5 MeV and Radius > 550 cm. Differential systematics are not shown.

These results show clearly that electron neutrinos are changing their flavor, violating a hypothesis test for no flavor change at greater than 7 σ. Previous measurements with pure heavy water had reached similar conclusions of flavor change at levels of 5.3 σ [3,4] (by consideration of the summed CC and NC and ES data) and 3.3 σ [2] (by comparison of the CC data with ES measurements by Super-Kamiokande). Oscillation purely to sterile neutrinos is also strongly disfavored and sub-dominant oscillations to sterile neutrinos are restricted by comparison of the active total flux of $^8$B solar neutrinos observed by the NC reaction with the $^8$B flux calculated with solar models. The measurements from the salt phase, combined with the results of other solar neutrino experiments indicate that the mixing is non-maximal (mixing angle less than π/4) with a confidence level corresponding to 5.5 standard deviations. The data for solar neutrinos have been analyzed by many authors. Figure 4 shows the results for an analysis assuming $\nu_1$, $\nu_2$ mixing is dominant [5], following the SNO salt data. (Some analyses also allowed a very small region at about three orders of magnitude lower in $\Delta m^2$ at a 99.73 % confidence level.). The clear indication for matter enhancement of the oscillation via the MSW effect in the LMA region enables the sign of $\Delta m^2$ to be determined unambiguously and it is found that $m_2$ is greater than $m_1$.

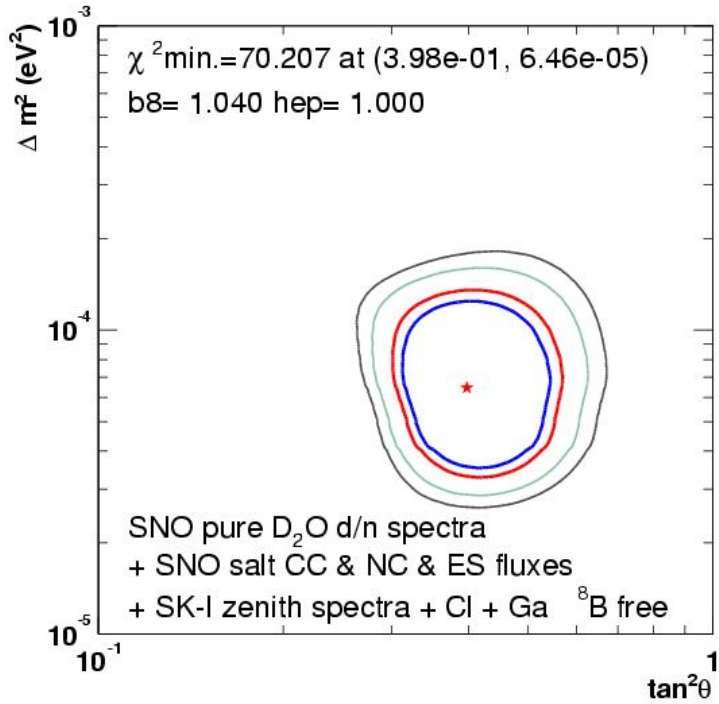

Figure 4. Global solar neutrino oscillation contours for two parameter mixing [5]. Data included in the fit is: $D_2O$ day and night spectra, salt CC, NC, ES fluxes, Super-Kamiokande spectral and day-night data, Cl, Ga fluxes . The best-fit point is $\Delta m^2 = 6.5 \times 10^{-5}$ eV$^2$, $\tan^2 \theta = 0.40$. The four contour lines correspond to 90, 95, 99, 99.73 % confidence limits.

## 3 Reactor Neutrino Measurements

For many years, reactor neutrino measurements sought evidence for flavor change of anti-neutrinos but no measurements showed clear effects. The distances from the reactors ranged from 10 m to 1 km. However, the results of the CHOOZ [17] and Palo Verde [18] measurements proved to have value in defining neutrino oscillation parameters. When combined with the finite oscillation effects observed by Super-Kamiokande for atmospheric neutrinos, they provide significant limits on the $\theta_{13}$ mixing angle, giving an upper limit of 0.066 for $\sin^2\theta_{13}$ at 3 $\sigma$. [19]

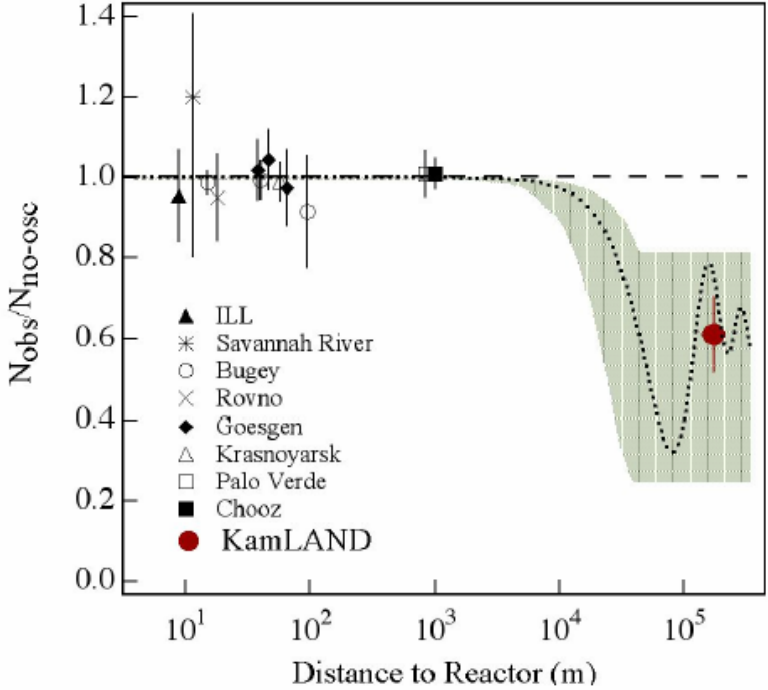

Figure 5. Summary of measurements of neutrino oscillation for reactor anti-neutrinos. See reference [7].

### 3.1 KamLAND Measurements for Reactor Anti-Neutrinos

As shown in Figure 5, the KamLAND experiment used anti-neutrinos from all the reactors in Japan and Korea (flux averaged distance of about 180 km) to observe [7] a finite disappearance effect for a set of oscillation parameters that corresponded closely to the LMA parameters for solar neutrinos shown in Figure 4.

Figure 6 shows the allowed ranges for the latest data from KamLAND [20]. As can be seen, assuming CPT invariance, there is excellent agreement between the allowed oscillation parameters for the solar neutrino measurements and the KamLAND reactor anti-neutrino results. If the results are combined, the allowed region shown in Figure 6 (b) is very small. The principal restriction on the $\theta_{12}$ axis comes from the solar (principally SNO) measurements and for the $\Delta m^2$ axis, from the KamLAND results. This arises because the $^8$B solar neutrinos in the LMA region are undergoing a resonant MSW transition and emerge from the Sun in essentially a pure $\nu_2$ state. Then, for small $\theta_{13}$, the electron neutrino survival probability (CC/NC fluxes for SNO) is almost exactly $\sin^2\theta_{12}$. On the other hand, the vacuum oscillation probability for the reactor anti-neutrinos $P(\nu_e \rightarrow \nu_e)$ is very sensitive to $\Delta m^2$, as shown in the equation above.

The agreement between the solar neutrino oscillation parameters and the reactor anti-neutrino oscillation parameters implies that MNSP mass oscillations are the dominant process for flavor change. Many other possible flavor changing processes are relegated to sub-dominant roles at most. The agreement provides a confirmation of CPT for neutrinos and if full CPT invariance is assumed, the agreement also confirms matter enhancement for the solar neutrinos and restricts the possibility of flavor change arising from Resonant Spin Flavor Precession [21] in the Sun. Sub-dominant transitions to sterile neutrinos are also restricted significantly [22]. The measurements to date of flavor change in solar, reactor and atmospheric neutrinos also place restrictions on possible processes such as neutrino decay, decoherence, violation of the equivalence principle, violation of Lorentz Invariance and flavor-changing neutral currents.

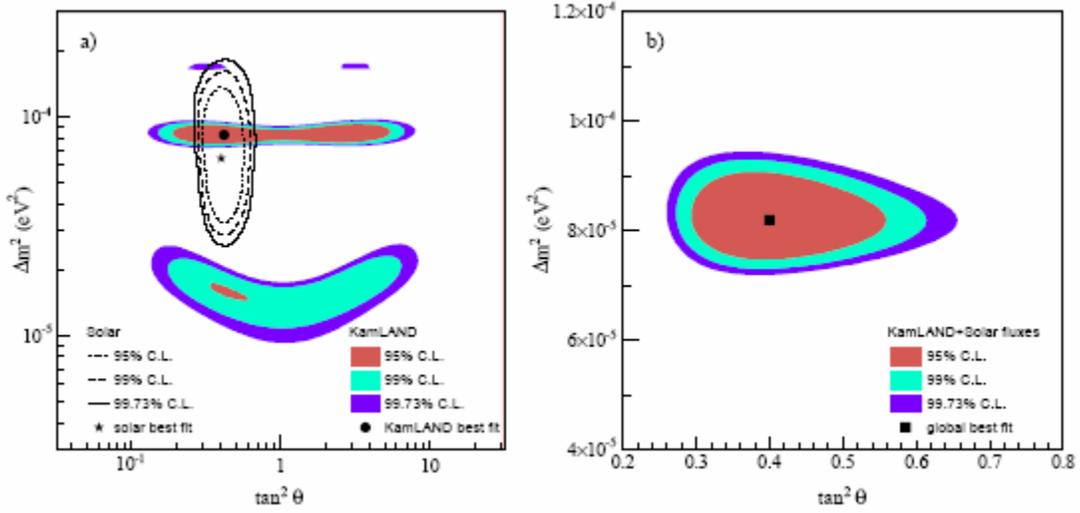

FIG. 6: (a) Allowed regions of neutrino oscillation parameters from KamLAND anti-neutrino data (shaded regions) and solar neutrino experiments (lines) [20]. (b) Result of a combined two-neutrino oscillation analysis of KamLAND and the observed solar neutrino fluxes under the assumption of CPT invariance. The best-fit point is:
$\Delta m_{12}^2 = 8.2^{+0.6}_{-0.5} \times 10^{-5} \, eV^2$ and $\tan^2 \vartheta_{12} = 0.40^{+0.09}_{-0.07}$.

## 4. Future Measurements

The set of measurements described above have provided a restricted range of values for $\theta_{12}$, $\theta_{13}$ and $\Delta m_{12}$. As will be described in the following paper, restricted ranges for $\theta_{23}$ and $\Delta m_{23}$ have also been obtained from measurements of atmospheric neutrinos. Future measurements are planned for all of these areas with the intention of improving the accuracy of these elements of the MNSP matrix and test theoretical models that extend the standard model in the neutrino sector. Another objective of future measurements is to test the completeness of the MNSP model as an explanation for flavor change and to seek or limit other sub-dominant processes. In the case of solar neutrino measurements, there is also the objective of understanding solar models more completely, including, for example, explicit measurement of neutrinos from the CNO cycle.

### 4.1 Future Measurements of Solar Neutrinos

The SAGE experiment continues to provide data on solar neutrino fluxes detected radiochemically with Ga, including the pp neutrino flux. The Super-Kamiokande has returned to operation with fewer photomultipliers after their accident in 2002. It is obtaining solar neutrino data with a slightly higher threshold and will shut down in 2005 to restore the complement of PMT's to the original number and return to the former threshold for solar neutrino detection. The SNO experiment has entered its third phase with an array of $^3$He-filled proportional detectors installed in the heavy water. In this phase, the NC and CC reactions will be detected independently, removing the correlations between them and providing a more accurate measure of the NC flux. This set of measurements will continue to improve the accuracy of the 1,2 mixing parameters. As discussed above, the improvements in the SNO data provides an almost direct determination of the $\theta_{12}$ parameter. In addition, the combination of new SNO with further KamLAND reactor anti-neutrino measurements will continue to restrict the range of $\theta_{13}$ for low values of $\Delta m_{23}^2$.

Future measurements of low energy solar neutrinos will be directed at real-time individual observation of pp, pep, $^7$Be and CNO neutrinos. In the case of pp neutrinos there are experiments aiming at CC measurements and others aimed at ES measurements from electrons, so that the combination can be used to make a measurement of neutrino oscillations at this energy without reference to solar models (by using the small sensitivity of the ES reaction to NC processes). However, the uncertainties in solar model fluxes for the pp and pep reactions are on the order of a few percent [6] as they are strongly constrained by solar luminosity. Therefore CC measurements alone can provide an accurate oscillation measurement for these neutrino fluxes. It is interesting to compare the survival probability for pp and pep neutrinos with those measured for $^8$B solar neutrinos, for the following reason. The $^8$B neutrinos undergo an MSW resonant process in the Sun, whereas at the pp or pep energies, the oscillation is essentially pure vacuum oscillation. Therefore, in the two neutrino approximation, the survival for the $^8$B neutrinos is $\sin^2 \theta_{12}$ and for the pp neutrinos it is $1 - 1/2 \sin^2 2\theta_{12}$. A comparison of these two determinations can test these models in details as well as providing a more accurate determination of $\theta_{12}$ and further restrictions on sterile neutrino contributions, especially sterile neutrino parameters that would result in spectral distortions in the low energy region [23].

There are many future measurements under development that would observe lower energy solar neutrinos by the ES and CC reactions. The ES measurements include liquid scintillator detectors for $^7$Be neutrinos: Borexino [24] that should begin data taking within the next year and KamLAND, for which background improvements should permit solar neutrino data in 2007 [25]. For pp neutrinos, liquid scintillator is limited by $^{14}$C background and so liquid noble gasses such as He (Heron) [26], Ne (CLEAN) [27] and Xe (XMASS) [28] are being developed. For CC measurements, Moon [29] with Mo and LENS [30] with In are under development.

**4.2 Future Reactor Neutrino Measurements**

Future reactor neutrino measurements are primarily focused on the determination of the mixing parameter $\theta_{13}$ by searching with a baseline corresponding to the characteristic length for $\Delta m_{atmospheric}$ and reactor anti-neutrino energies. The relevant survival probability for reactor anti-neutrinos is:

$$P(\bar{\nu}_e \to \bar{\nu}_e) \cong 1 - \sin^2 2\theta_{13} \sin^2\left(\frac{\Delta m^2_{atm} L}{4E}\right) - \cos^2 \theta_{13} \sin^2 2\theta_{12} \sin^2\left(\frac{\Delta m^2_{12} L}{4E}\right).$$

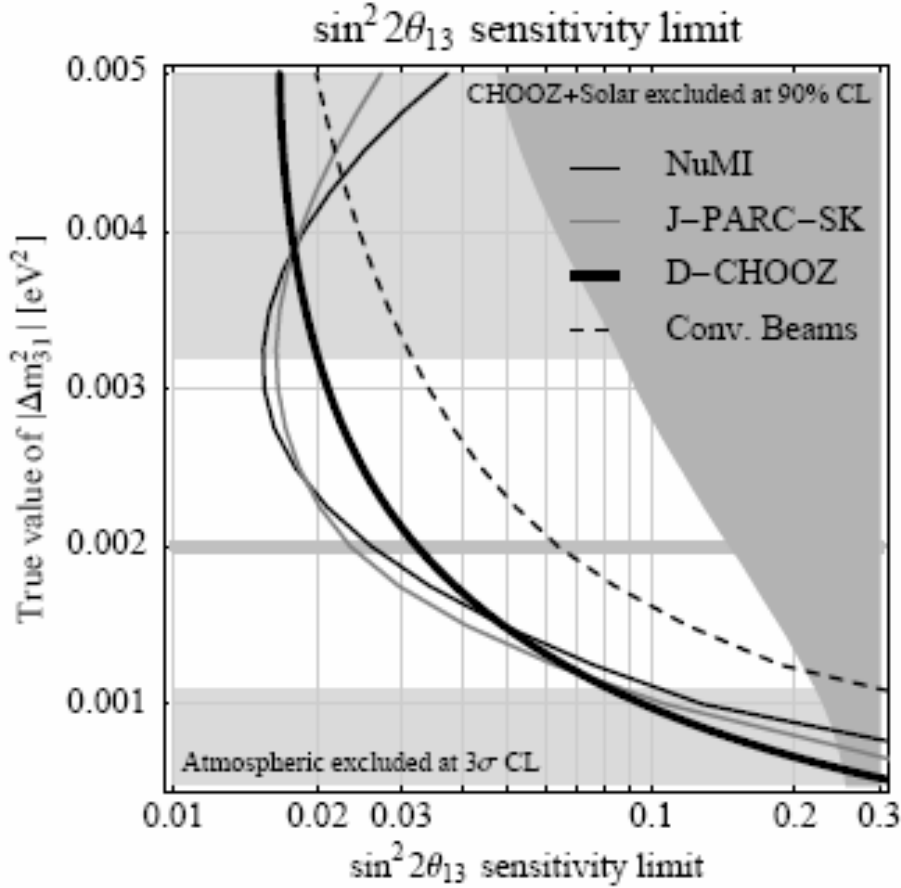

Figure 7. Sensitivity for the observation of the mixing parameter $\theta_{13}$ in various next-generation reactor and accelerator measurements.

By choosing the L to maximize the first term, sensitivity to $\theta_{13}$ is obtained.

Figure 7 shows the sensitivity for one of the proposed experiments (Double-CHOOZ) [31] at the CHOOZ reactor in France. Other experiments with comparable sensitivity are in the planning stages for reactors in the United States, China, Brazil and Russia.[32] A successful determination of $\theta_{13}$ will be very important for the design of major long baseline accelerator experiments seeking the observation of $\delta_{CP}$.

## 5. Conclusions

Measurements of solar neutrinos and reactor anti-neutrinos have clearly observed neutrino flavor change, via appearance and disappearance measurements. The combination of these observations has defined the dominant flavor change mechanism to be oscillation of massive neutrinos via the MNSP formalism. The mixing parameters for mass 1 and 2 are accurately defined by the combination of these measurements. Other mechanisms are sub-dominant if they occur, and the levels of contribution are limited, including oscillation to sterile species. The MSW effect provides matter enhancement of neutrino oscillation in the Sun and defines a normal hierarchy for mass 1 and 2. The total flux of $^8$B solar neutrinos is observed to be very close to the predictions of the Standard Solar Model.

Future measurements will improve the accuracy of oscillation parameters significantly and thereby provide significant constraints on theoretical models that extend the Standard Model for elementary particles.